\documentclass[12pt, reqno]{article}

\usepackage{rotating} 
\usepackage{amssymb} 
\usepackage{amsmath}
\usepackage{amsxtra} 
\usepackage{geometry} 
\usepackage{appendix}

\makeatletter
\@addtoreset{chapter}{part}
\@addtoreset{@ppsaveapp}{part}
\makeatother

\usepackage[longnamesfirst]{natbib} 

\usepackage{lscape} 

\usepackage{color}  

\usepackage{float} 

\usepackage{standalone} 

\definecolor{ChadDarkBlue}{rgb}{.1,0,.2}
\definecolor{ChadBlue}{rgb}{.1,.1,.5}
\definecolor{ChadRoyal}{rgb}{.2,.2,.8}
\definecolor{ChadGreen}{rgb}{0,.4,0}    
\definecolor{ChadRed}{rgb}{.5,0,.5}  

\usepackage[format=plain, 
font=sc,  
labelfont=bf, 
skip=10pt, 
justification=centering
]{caption}

\usepackage{hyperref}
\hypersetup{
	colorlinks=true,        
	linkcolor=ChadRed,
	urlcolor=ChadGreen,
	citecolor=ChadRed,
	pdftitle={CollateralValue},
	pdfauthor={ },
	pdfstartview={FitV},
	pdfpagemode={UseNone},
	pdfnewwindow=true,      
}

\usepackage{setspace}

\newcommand{\gaz}{>}

\renewcommand{\epsilon}{\varepsilon}

\usepackage{booktabs,caption}
\usepackage[flushleft]{threeparttable}

\title{
Does Bankruptcy Protection Affect Asset Prices?
\\
Evidence from changes in Homestead Exemptions\thanks{
	All errors are our own. 
}
	}
\author{
	{Yildiray Yildirim and Albert Alex Zevelev}\thanks{
		Email: 
		\href{mailto:Yildiray.Yildirim@baruch.cuny.edu}{Yildiray.Yildirim@baruch.cuny.edu},
		\href{mailto:Albert.Zevelev@baruch.cuny.edu}{Albert.Zevelev@baruch.cuny.edu}
	}
	\\ \\
	Baruch 
	\\
	October 15, 2019
}
\date{} 

\usepackage{graphicx} 
\begin{document}
\maketitle
\begin{abstract}
Does the ability to protect an asset from unsecured creditors affect its price?
This paper identifies the impact of bankruptcy protection on house prices using 139 changes in homestead exemptions.
Large increases in the homestead exemption raised house prices 3\% before 2005. 
Smaller exemption increases, to adjust  for inflation, did not affect house prices. 
The effect disappeared after BAPCPA, a 2005 federal law designed to prevent bankruptcy abuse.
The effect was bigger in inelastic 
locations. 
\end{abstract}
\onehalfspacing
\newpage
%
\section{Introduction 
}
\label{Introduction}
Bankruptcy is one of the largest social insurance programs in the US. 
Americans discharge more debt (formally and informally) than all unemployment benefits
combined.\footnote{\cite{lefgren2010, auclert2019, pattison2019asset}.}
The sharp rise in personal bankruptcy from .3\% of households annually in the 1980s 
to 1.5\% in the early 2000s raised concern about strategic behavior and motivated 
the 2005 Bankruptcy Abuse Prevention and Consumer Protection Act 
(BAPCPA).\footnote{\cite{gross2019, albanesi2018}.}
The bill raised the costs and reduced the benefits of filing for bankruptcy. 

A large literature studies strategic bankruptcy. 
\cite{pattison2019asset} found that rises in homestead exemptions
are followed by a rise in chapter 7 filings by debtors with home equity.
\cite{helland2016self} found that in states with unlimited homestead exemptions 
physicians invest 13\% more in their homes 
compared to other professionals with similar income and demographics. 
Additionally, the response of physicians to unlimited
homestead exemptions is larger in areas with higher liability risk.
While the literature has found that households prefer to own assets which give them 
greater protection, we study whether the market price of these assets reflects this protection. 

There are several advantages of using house prices to detect strategic behavior. 
First, studies that use formal bankruptcy filings do not capture a lot of strategic behavior
since the majority  of defaulting consumers  do not file for bankruptcy,   
and   most   debt   collection  takes   place outside  of the courtroom (\cite{dawsey2013non}). 
However, the settlement negotiated between creditors and debtors outside the courtroom 
is influenced by local exemption laws
under the ``threat-point" of bankruptcy.\footnote{\cite{mahoney2015, skeel2003, pattison2017consumption}.}
Second, changes in house prices in locations with a rise in homestead exemptions can help quantify 
forward looking strategic behavior via demand for assets that provide this implicit insurance.

This study uses annual house price data in 55,316 census tracts from the FHFA
combined with 139 changes in homestead exemptions between 1990-2017, 
collected from other authors,\footnote{We thank Mariela Dal Borgo, Richard Hynes, 
Paul Goldsmith-Pinkham, and Jeffrey Traczynski for sharing their data.} 
legal guide books (\cite{elias1989file}), and legal statutes.\footnote{We thank Albert Levi for helping us 
search statutes.} 
The identifying assumption is that changes in homestead exemptions
are uncorrelated with unobserved determinants of house prices. 
This assumption has been confirmed by a vast legal and economic 
literature (see \hyperref[Institution]{Section \ref*{Institution}}) and by our own falsification tests. 

We find that the unconditional average rise 
in homestead exemptions raises real house prices $0.73\%$, 
an effect which is positive, statistically significant, and small. 
However, when the sample is restricted to Pre-BAPCPA observations (when bankruptcy was cheaper and easier), 
the treatment effect rises to $1.07\%$, 
and Post-BAPCPA (when bankruptcy was more expensive and less beneficial), 
the effect is no longer statistically significant. 
Next, when the sample is restricted to ``large" changes in homestead exemptions 
(defined to be changes greater than or equal to $\$50,000$)
the effect rises to $1.82\%$ and is not statistically significant for small changes. 
Finally, big changes Pre-BAPCPA raise house prices $3.04\%$, whereas small changes 
Pre-BAPCPA and all changes Post-BAPCPA have no statistically significant effect. 
Together these estimates reveal evidence of strategic behavior by households to protect 
their assets before BAPCPA (when bankruptcy was cheaper, easier, and more financially beneficial). 
These estimates also indicate that BAPCPA achieved its stated goal of reducing 
bankruptcy abuse. 

The main results are validated by dynamic regressions which find parallel pre-trends. 
A heterogeneity analysis finds no effect for reductions in the homestead 
exemption\footnote{In 1993, Minnesota reduced its homestead exemption from unlimited to \$200,000. 
This is the only negative change in the homestead exemption in our sample.} on house prices indicating an asymmetric effect.
In addition, census-tracts in Metropolitan Statistical Areas with relatively inelastic
housing supply had experienced bigger effects, consistent with standard theory.  
Moreover, tracts in counties with higher pre-treatment unemployment rates 
had smaller effects. This doesn't mean that households in counties with higher unemployment 
rates don't value bankruptcy protection, but rather that the types of households 
who strategically protect their assets from creditors tend to be more prosperous. 

Finally, falsification tests find that changes in the homestead exemption 
don't affect levels and changes in 
unemployment rates,
income per capita, 
and single family building permits. 
There is a small drop in population levels and homeownership rates.
The drop in homeownership rates suggests that 
the rise in house prices caused by wealthier strategic households 
reduced housing affordability.



\section{Institutional Setting}
\label{Institution}
This section describes the institutional setting and relevant details of the bankruptcy system 
in the United States. 
If a debtor defaults on secured debt the creditor can seize the collateral.  
If there is a deficiency\footnote{A deficiency is when the collateral is
worth less than the debt balance.} and recourse is possible\footnote{Recourse is possible when it is  
legal and not restricted in the original debt contract.} then the debtor is personally liable for the 
deficiency.\footnote{
The deficiency on the secured debt becomes unsecured debt.} 
If a debtor defaults on 
unsecured debt (e.g. credit card debt, legal judgment, student loan debt)
the creditor can seize the debtor's non-exempt assets and income. 
However, asset and income exemptions (i.e. wage garnishment limits) depend on local and federal laws 
which vary over time and space. 

For example, suppose Ann is a homeowner with a house worth \$200,000 
and a mortgage balance of \$80,000. Ann has $Heq_{i,t}=\$120,000$ in home equity. 
Ann lives in a state with a homestead exemption of $H_{i,t}=\$50,000$.  
This means that Ann currently has $Heq_{i,t} - H_{i,t}=\$70,000$ in unprotected 
home equity.
Suppose Ann defaults on \$90,000 in credit card debt. 
If the unsecured creditor forces a foreclosure and \$200,000 is recovered then
the mortgage lender will be paid \$80,000 first, 
next Ann can keep $H_{i,t}=\$50,000$ of her exempt home equity, 
the unsecured lender will receive the remaining \$70,000.
In practice the unsecured lender may not force a foreclosure, 
but instead place a \$90,000 lien on the home. 
The lien-holder will be repaid the debt plus fees and interest at a future date.  
Either way, the creditor and debtor both understand that Ann currently has 
\$70,000 of unprotected (seizable) home equity.

\begin{figure}[!ht]
	\centering
	\small \caption{Homestead Exemption Levels} 
	\label{F2_HexLevelb}
	\includegraphics[scale=.666]{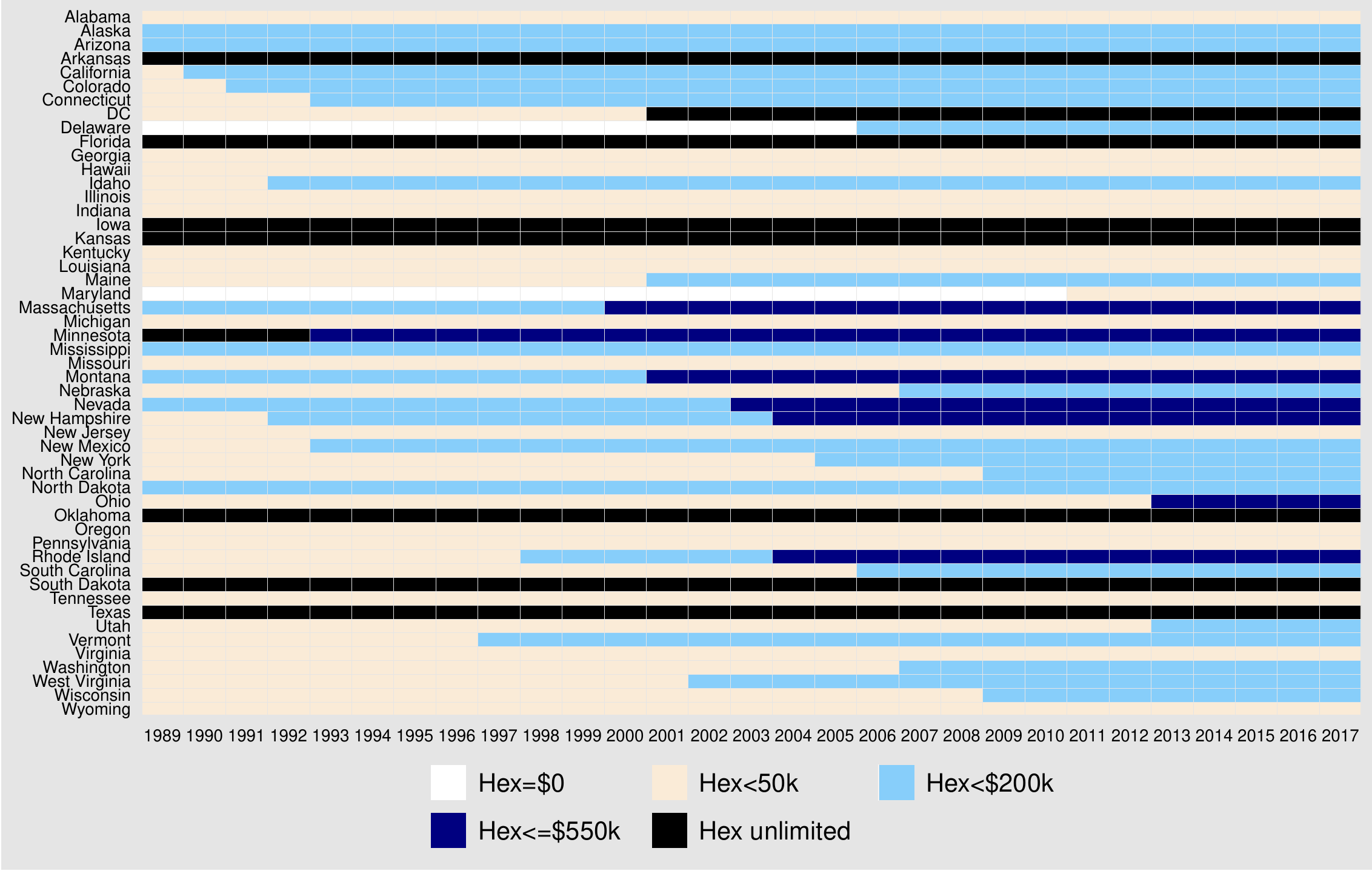}
	\begin{minipage}{0.7\textwidth}
		{\footnotesize 
			\textit{Note}. This figure presents all homestead exemption levels from 
			1989-2017.
The data was collected from other authors, bankruptcy guidebooks, and statutes as described in the paper.			  
			\par}
	\end{minipage}
\end{figure}

Homestead exemption laws vary over time and space (\hyperref[F2_HexLevelb]{Figure \ref*{F2_HexLevelb}}).
For example, if Ann lived in Delaware before 2005 or Maryland before 2010, 
she would have \$0 in protected home equity, her creditors would know this
and this would affect the settlement reached if it was negotiated out of court. 
On the other hand, if Ann lived in one of the eight states with unlimited homestead exemption, 
or any state with a homestead exemption over $Heq_{i,t}=\$120,000$ all her home equity would 
be protected, and her creditors would know this as well. 

The homestead exemption has been studied extensively in the history, legal, and economics literature. 
The first homestead exemption 
was 
incorporated into statutory and constitutional law  in the Republic of Texas in 1839 (\cite{london1954}).
Following the example of Texas, Mississippi passed the second homestead exemption law in 1841. 
Over the next few years, states throughout the US enacted homestead exemption laws. 
The literature\footnote{\cite{goodman1993, skeel2003}.} has argued that changes in homestead exemptions arise from a legislative process 
that depend on idiosyncratic historic 
and geographic\footnote{By 1860, there were seven states 
with homestead exemptions written into the constitution (in addition to statutory exemption). 
Six out of seven of these states were in the west. See the 
Turner hypothesis (\cite{london1954}).} factors, a process that does not depend 
on states' economic conditions.


In addition to the history and legal literature, an empirical literature has also studied 
determinants of changes in bankruptcy protection.\footnote{\cite{pattison2019asset, severino2017}.} 
\cite{severino2017} find that lagged changes in house prices, medical expenditure, 
unemployment rates, state GDP, bankruptcy filings, share of democrats, and income 
do not predict changes in homestead exemptions. 
We find further evidence that economic variables don't predict changes in homestead exemptions, 
over a longer sample.
Other bankruptcy protection laws, including  
statutes of limitations on debt, Tenancy by Entirety laws,\footnote{\cite{traczynski2019}.} and 
wage garnishment restrictions were stable over our sample period.

\section{Data}
\label{Empirical}
\label{Data}
This paper estimates the impact of changes in homestead exemption laws on changes in house prices. 
The main outcome variable, real house price growth, 
is measured using the recently available Federal Housing Finance Agency
\href{http://www.fhfa.gov/DataTools/Downloads/pages/house-price-index-datasets.aspx}
{(FHFA)} census tract data. 
This dataset contains 55,316 tracts in the US.
Like the S\&P/CoreLogic/Case-Shiller home price indices, the FHFA series 
corrects for the changing quality of houses being
sold at any point in time by estimating price changes with repeat-sales.
The dataset only includes tracts and years with enough repeat-sales 
to construct the index.\footnote{For details about this new dataset see \cite{Bogin}.}

%

The main treatment variable is the homestead exemption in location $i$ in year $t$ denoted 
$H_{i,t}$.\footnote{This data was collected from other authors (Mariela Dal Borgo, Richard Hynes, Paul Goldsmith-Pinkham, and Jeffrey Traczynski), legal guide books (\cite{elias1989file}), and legal statutes.} 
We define the homestead exemption as the maximum home equity that is protected:
\begin{align}
H_{i,t} & \equiv 
\max\left\{
H_{i,t}^{\text{Local, Married}},
1_{i,t}^{\text{Fed}} \text{Fed}_{t}^{\text{Married}} 
\right\}
\end{align}
where $H_{i,t}^{\text{Local, Married}}$ is the local homestead exemption for married households, 
$1_{i,t}^{\text{Fed}}$ is an indicator variable equal to one if local laws allow households 
to use the federal exemption,
and $\text{Fed}_{t}^{\text{Married}}$ is the federal homestead exemption level for married households in that year. 
Hence, $H_{i,t}$ is the maximum amount of home equity a household can legally protect
in a given location at a given time. 
The non-homestead exemption denoted $NH_{i,t}$ is the sum of the vehicle and wildcard exemptions, 
and is computed the same way as $H_{i,t}$. The wildcard exemption lets the debtor choose which 
property to protect such as a vehicle, bank deposits, and art. 

Data used for controls, heterogeneity analysis, and other outcome variables
come from several different sources.
Supply elasticity data are available at the MSA level from \cite{Saiz2010}.
Employment data at the county level are  from the BLS. 
Income data at the county level are from the BEA. 
Population, single family building permits, and homeownership data are from the census.
The population and permit data are at the county level, whereas the homeownership rate data 
are at the MSA level. 
Median house price data at the zip code level and rent data at the MSA level are from Zillow. 
One must be careful in merging the datasets since the same zip code can be in more than one county. 
Each zip code is assigned to the county with the maximum allocation factor 
(e.g. if $75\%$ of zip code $z$ is in county $A$ and $25\%$ in county $B$, 
then zip code $z$ is assigned to county $A$).
US oil price data are from the EIA. 
US interest rates are constructed as in \cite{HMS} by correcting the 
10 year Treasury bond rate for inflation with the Livingston Survey.
Nominal variables are deflated using the CPI for all urban consumers from the BLS as in \cite{GGG}.

\section{Estimates}
\label{Estimates}

\subsection{Summary Statistics} 
\label{Main_summ}
\begin{figure}[!ht]
	\centering
	\small \caption{Bankruptcy Protection Laws} 
	\label{F1}
	\includegraphics[scale=1]{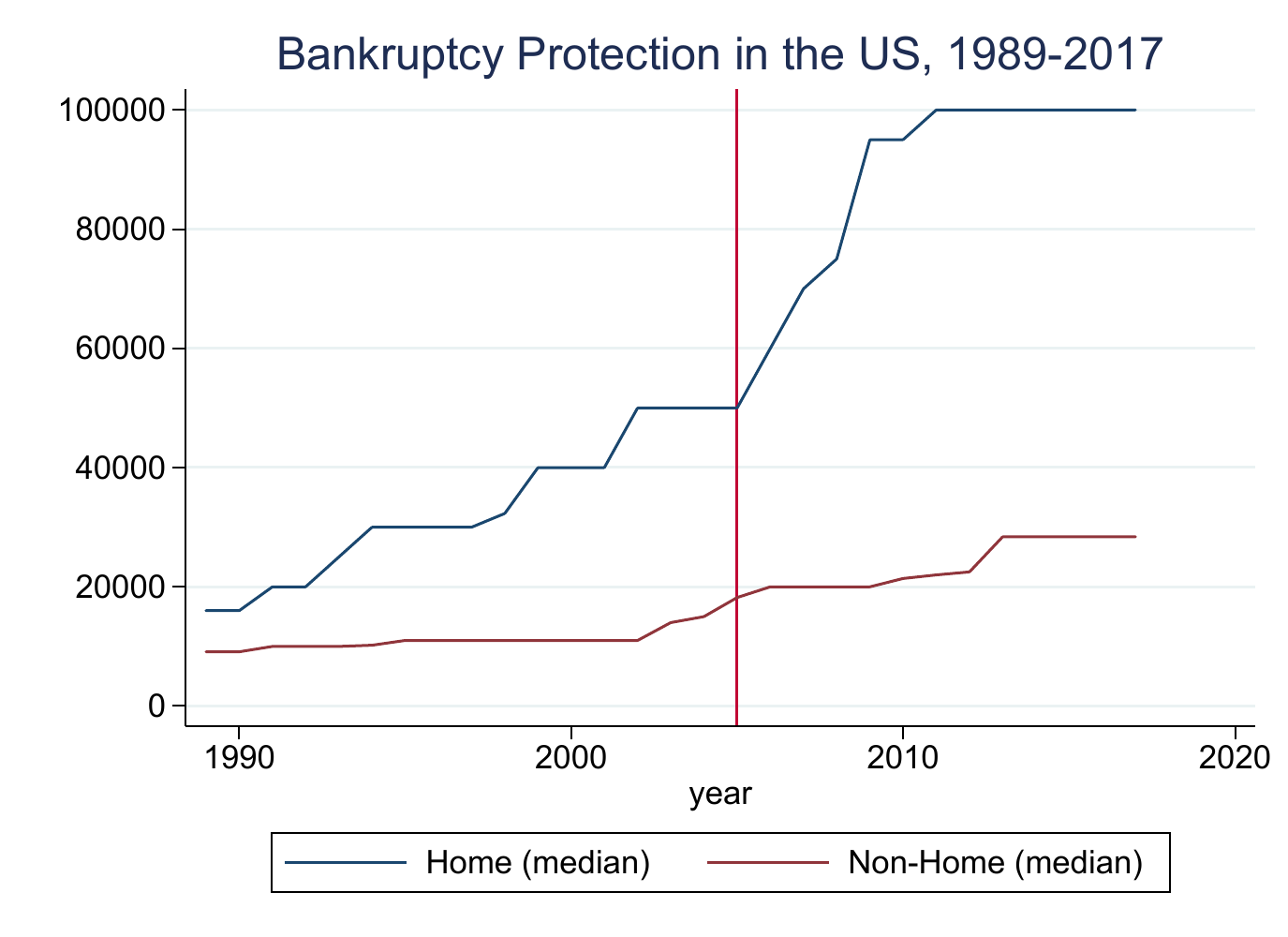}
	\begin{minipage}{0.7\textwidth}
		{\footnotesize 
			\textit{Note}. This figure plots median homestead and non-homestead exemptions 
			in the US between 1989-2017.
The data was collected from other authors, bankruptcy guidebooks, and statutes as described in the paper.			  			
			\par}
	\end{minipage}
\end{figure}

\hyperref[T1]{Table \ref*{T1}} presents descriptive statistics and
\hyperref[F1]{Figure \ref*{F1}} plots median homestead and 
non-homestead exemption
levels in the US 1989-2017.\footnote{Note our data on levels begins 1989 and our data on changes begins 1990.} 
A few stylized facts are immediately clear: 
1 both the homestead and non-homestead exemptions grew considerably over this sample, 
2 the homestead exemption grew a lot more and is currently much higher than the non-homestead exemption, 
3 changes in the homestead exemption are less frequent, occurring 9.73\% of the time
whereas changes in the non-homestead exemption occur 12.68\% of the time. 
Negative changes in bankruptcy protection laws are rare. 
The only reduction in the homestead exemption in our sample was in Minnesota in 1993, 
which reduced the homestead exemption from unlimited to \$200,000. 
The only reduction in the non-homestead exemption in our sample was in Louisiana in 2003,   
which reduced the motor vehicle exemption from \$24,000 to \$15,000.  

\subsection{Impact of Changes in Homestead Exemptions on House Prices}
\label{Main}
This section presents the main results, the impact of changes in homestead exemptions on real 
house price growth. We estimate:
\begin{align*}
y_{i,t} 
&=
\textcolor{ChadRed}{\beta_{H}}  1\left\{ \Delta H_{i,t}>0  \right\} 
+ g(X_{i,t})
+ u_{i,t}
\tag{Static} 
\\
y_{i,t} 
&=
\sum_{\underset{k\neq -1}{k=-3}}^{3}
\textcolor{ChadRed}{\eta_{k}}  1\left\{ \Delta H_{i,t}>0 \right\} 
+g(X_{i,t})
+u_{i,t}
\tag{Dynamic}
\end{align*}
where the main outcome variable $y_{i,t}$ is real house price growth in tract $i$ in year $t$, 
the main treatment variable $1\left\{ \Delta H_{i,t}>0 \right\}$ is an indicator for 
years when tract $i$ experienced a change in the homestead exemption, 
and $X_{i,t}$ are controls including tract and year fixed effects, the unemployment rate, population, 
and income per capita. 
The dynamic regression is used for validation,
to check whether pre-trends in the static model are parallel,  
and to investigate persistence in the treatment. 
In addition to using an indicator variable for the main treatment, we re-estimate the main equations 
using a continuous variable equal to change in homestead exemptions in the appendix. 

\hyperref[US_T1]{Table \ref*{US_T1}} presents the main results.
Column 1 finds that an (unconditional) average rise 
in homestead exemptions raises real house prices $0.73\%$, 
an effect which is positive, statistically significant, and small. 
However, when the sample is restricted to Pre-BAPCPA observations 
($t\leq 2005$ when bankruptcy was cheaper and easier), 
the treatment effect rises to $1.07\%$, 
and Post-BAPCPA ($t>2005$ when bankruptcy was more expensive and less beneficial), 
the effect is no longer statistically significant. 
Next, when the sample is restricted to ``large" changes in homestead exemptions 
(defined to be changes greater than or equal to $\$50,000$)
the effect rises to $1.82\%$ and is not statistically significant for small changes mostly due to inflation adjustment. 
Finally, big changes Pre-BAPCPA raise house prices $3.04\%$, whereas small changes 
Pre-BAPCPA and all changes Post-BAPCPA have no statistically significant effect. 
Together these estimates reveal evidence of strategic behavior by households to protect 
their assets before BAPCPA (when bankruptcy was cheaper, easier, and more financially beneficial). 
These estimates also indicate that BAPCPA achieved its stated goal of reducing 
bankruptcy abuse. 

\begin{figure}[!ht]
	\centering
	\small \caption{Impact of Changes in Homestead Exemptions on Real House Price Growth} 
	\label{FDyncov}
	\includegraphics[scale=1.1]{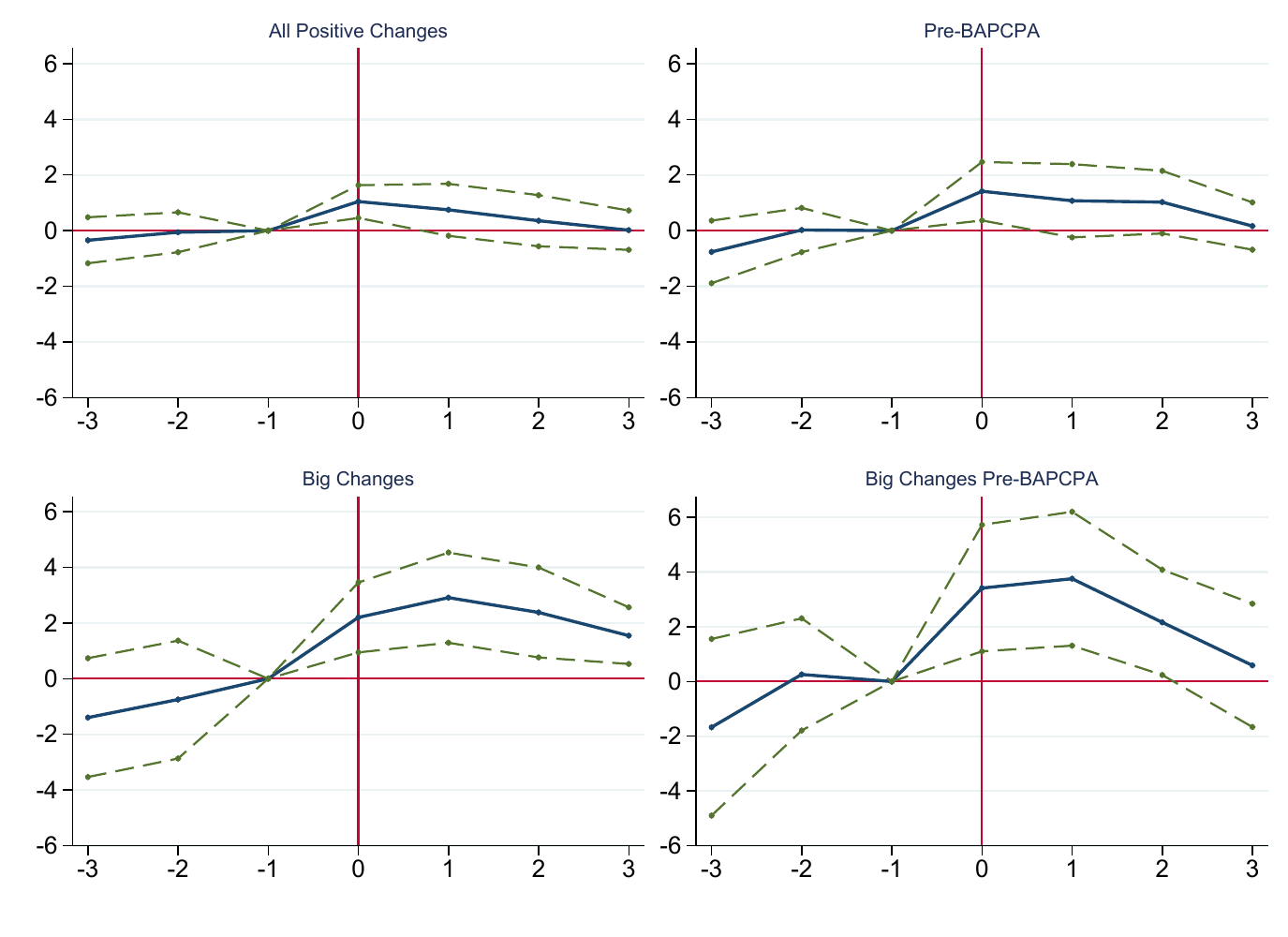}
	\begin{minipage}{0.7\textwidth}
		{\footnotesize 
			\textit{Note}. This figure plots point estimates 
$\textcolor{ChadRed}{\hat{\eta}_{k}}$ and 95\% confidence intervals from 
the dynamic regression in \hyperref[T2_Dyncov]{Table \ref*{T2_Dyncov}}.
There is a vertical red line in the year of the law change.
House price data is at the census-tract year level from the FHFA, deflated by the CPI-U.
The bankruptcy law data was collected from other authors, bankruptcy guidebooks, and statutes as described in the paper.			  
			\par}
	\end{minipage}
\end{figure}

Next, estimates from corresponding dynamic regressions are 
presented in \hyperref[T2_Dyncov]{Table \ref*{T2_Dyncov}}
and plotted in \hyperref[FDyncov]{Figure \ref*{FDyncov}}. 
The first pattern we observe is that pre-trends are parallel in all four specifications, 
which is encouraging.
Second, we see in columns 3 and 4, 
as the estimates become larger they also become more persistent.

Next \hyperref[US_T1rob]{Table \ref*{US_T1rob}}
repeats the same analysis as above except using continuous changes in the homestead 
exemption, as opposed to an indicator, as the main treatment variable $\Delta H_{i,t}$
divided by the average big (over \$50,000) change which is \$115,169.33. 
The results are similar to the results estimated using indicator variables for treatments above. 
\hyperref[US_T1Dynrob]{Table \ref*{US_T1Dynrob}} presents corresponding dynamic estimates
which also show parallel pre-trends.


\subsection{Treatment Effect heterogeneity}
\label{ResultsHTE}

This section investigates heterogeneity in the treatment effect -- that is, 
whether the treatments had different impacts in different locations. 
To study the sensitivity of the effect to  various observable measures of heterogeneity $G_{i}$, 
this paper estimates:
\begin{align*}
y_{i,t} 
&
=
\textcolor{ChadRed}{\beta_{H,0}}  1\left\{ \Delta H_{i,t}>0  \right\} 
+
\textcolor{ChadRed}{\beta_{H}} 1\left\{ \Delta H_{i,t}>0  \right\}  \times G_{i}
+
g(X_{i,t})
+
\epsilon_{i,t}
\tag{DDD} 
\end{align*} 
In this specification the average treatment effect (ATE) is an affine function of $G_{i}$ 
\begin{align*}
\text{ATE}\left(G_{i}\right) 
&= 
\textcolor{ChadRed}{\beta_{G,0}} + \textcolor{ChadRed}{\beta_{G}}  G_{i}  
\end{align*}
The coefficient $\textcolor{ChadRed}{\beta_{G,0}}$ 
is the estimated average treatment effect if $G_{i} =0$
and  
$\textcolor{ChadRed}{\beta_{G}} 
= \frac{\partial \text{ATE}\left(G_{i}\right)  }{ \partial G_{i}}  $ is 
the sensitivity of the average treatment effect to a rise in $G_{i}$.

\begin{figure}[!ht]
	\centering
	\small \caption{The impact of a rise in demand on house prices
		in cities with different supply elasticities} 
	\label{ElasticSupply}
	\includegraphics
	[scale=1]
	{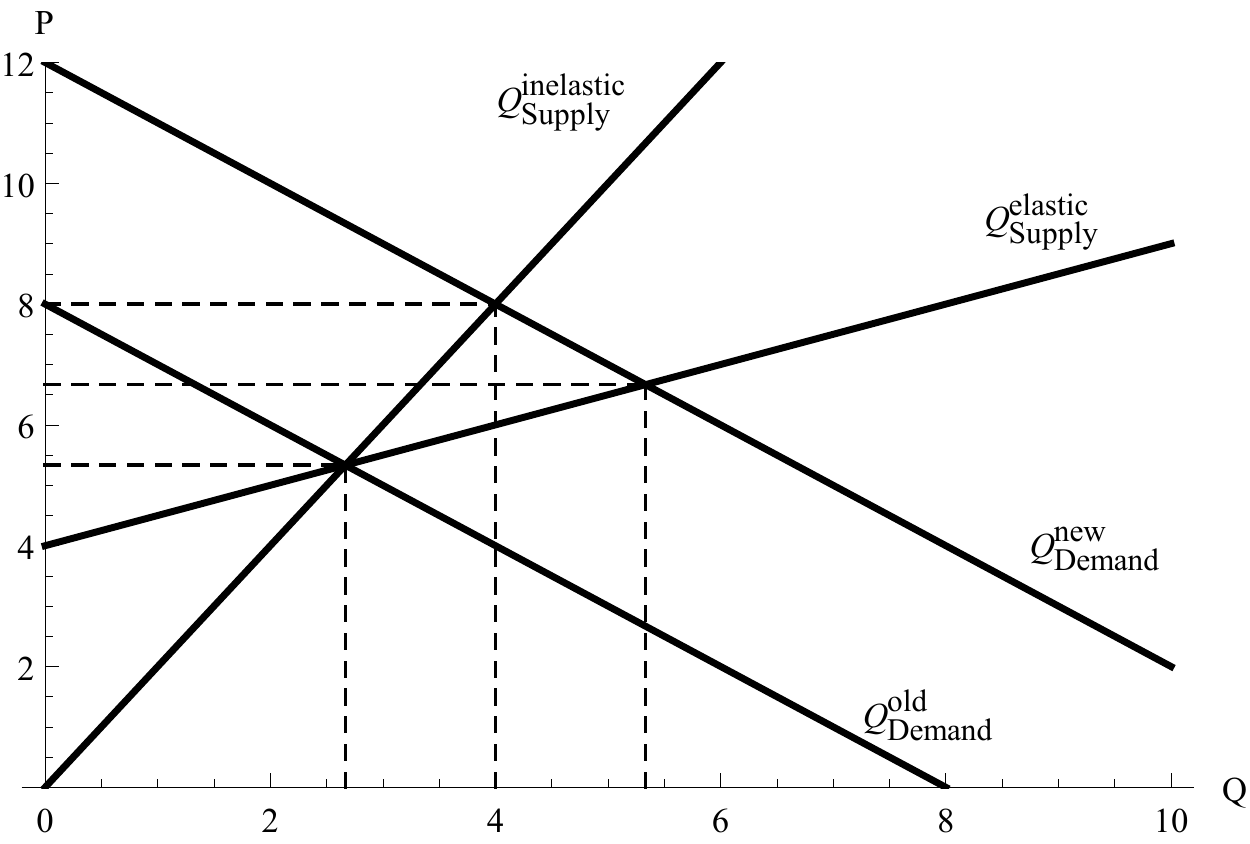}
	\begin{minipage}{0.7\textwidth}
		{\footnotesize 
			\textit{Note}. This figure compares the impact of a rise in
			housing demand on house prices in 
			two cities with different supply elasticities.
			Price (P) is on the vertical axis and quantity (Q) in on the horizontal axis. 
			Initially, the price of housing is the same in both cities. 
			A rise in demand 
			causes prices
			to rise more in the relatively inelastic city. 
			\par}
	\end{minipage}
\end{figure}

For example, theory predicts that a rise in demand should have 
a smaller impact on house prices in elastically supplied locations 
where it is easier to build real estate (\hyperref[ElasticSupply]{Figure \ref*{ElasticSupply}}). 
This corresponds to the hypothesis 
$\textcolor{ChadRed}{\beta_{Elasticity}} <0$.
The coefficient $\textcolor{ChadRed}{\beta_{Elasticity,0}}$ is the estimated impact of 
the law change on prices in a hypothetical location where the asset (housing) 
is in perfectly inelastic supply.

\hyperref[THTE3]{Table \ref*{THTE3}} investigates treatment effect heterogeneity 
in positive versus negative changes in the homestead exemption, supply elasticity, 
\textit{pre-treatment} unemployment rates, population, real income per capita,
median real house value, home ownership rates, and single family building permits.
Pre-treatment variables are set equal to their value in the year before the treatment to make 
sure they are unaffected by the treatment. 
The negative change in homestead exemption in Minnesota in 1993, had a modest effect of -0.31\%,
however it is not statistically significant. 
Consistent with theory (\hyperref[ElasticSupply]{Figure \ref*{ElasticSupply}}) supply elasticity
attenuates the treatment effect. A 1\% higher supply elasticity corresponds to 0.62\% smaller effect. 
The implied treatment effect for a hypothetical city with perfectly inelastic housing supply 
is 1.68\%. 
The only other significant source of heterogeneity is the pre-treatment unemployment rate. 
Locations with higher pre-treatment unemployment rates had much smaller effects. 
This doesn't necessarily mean that households in these locations don't value bankruptcy protection, 
it only suggests that these types of households are less likely to be strategic in protecting their assets
(possibly because they don't have the same access to financial advisors and tax attorneys as households in wealthier areas).

\subsection{Channels} 
\label{Main_channels}
This section explores the mechanism through which changes in homestead exemptions affect 
house prices. 
On the one hand, the effect could be driven by a rise in demand for bankruptcy protection. 
On the other hand, the rise in bankruptcy protection can increase entrepreneurship 
affecting local demand. 
\hyperref[THTEY]{Table \ref*{THTEY}} examines the impact of changes in the homestead exemption 
on alternative outcome variables including levels and first differences of:
unemployment rates, population, real income per capita, single family building permits, 
and home ownership rates. 
A rise in homestead exemptions has a small negative effect on population  levels (but not changes), 
and on home ownership rates (but not changes).
Levels and changes in unemployment rates and real income per capita are not affected. 
Together these results indicate that changes in homestead exemptions likely have 
a very small, if any, impact on local demand.


\subsection{Predictors of Law Changes} 
\hyperref[THTEp]{Table \ref*{THTEp}} examines determinants of changes in homestead exemptions  
using various  predictors including lagged:
real house price growth, unemployment rates, population, real income per capita, homeownership rates, 
homestead exemption levels, and non-homestead exemption levels. 
The only consistent, statistically significant predictor of changes in homestead exemptions is
the lagged homestead exemption level. Locations with high (or unlimited) homestead 
exemptions are less likely to raise them, compared to locations with low (or zero) exemptions
that want to catch up.


\section{Conclusion}
\label{Conclusion}
A large body of literature studies the impact of bankruptcy protection on 
bankruptcy filings. 
These studies do not capture a lot of strategic behavior
since the majority  of defaulting consumers  do not file for bankruptcy,   
and   most   debt   collection  takes   place outside  of the courtroom (\cite{dawsey2013non}). 
In contrast, we use  house prices to quantify demand for bankruptcy protection. 

We find that the average rise in homestead exemptions raises real house prices $0.73\%$, 
an effect which is positive, statistically significant, and small. 
However, when the sample is restricted to Pre-BAPCPA observations 
($t\leq 2005$ when bankruptcy was cheaper and easier), 
the treatment effect rises to $1.07\%$, 
and Post-BAPCPA ($t>2005$ when bankruptcy was more expensive and less beneficial), 
the effect is no longer statistically significant. 
Next, when the sample is restricted to ``large" changes in homestead exemptions 
(defined to be changes greater than or equal to $\$50,000$)
the effect rises to $1.82\%$ and is not statistically significant for small changes, mostly due to inflation adjustment. 
Big changes Pre-BAPCPA raise house prices $3.04\%$, whereas small changes 
Pre-BAPCPA and all changes Post-BAPCPA have no statistically significant effect. 
Together these estimates reveal evidence of strategic behavior by households to protect 
their assets before BAPCPA (when bankruptcy was cheaper, easier, and more financially beneficial). 
These estimates also indicate that BAPCPA achieved its stated goal of reducing 
bankruptcy abuse.

\newpage
\section*{}
\bibliographystyle{jf}
\bibliography{AZ_master}

%

\newpage
\begin{appendices}
\section{Appendix: Figures}
\subsection{Bankruptcy Protection Laws}

\newpage
\subsection{Homestead Exemption Law Changes}
\begin{figure}[!ht]
	\centering
	\small \caption{Homestead Exemption Law Changes} 
	\label{F2_AllChangeHex}
	\includegraphics[scale=.666]{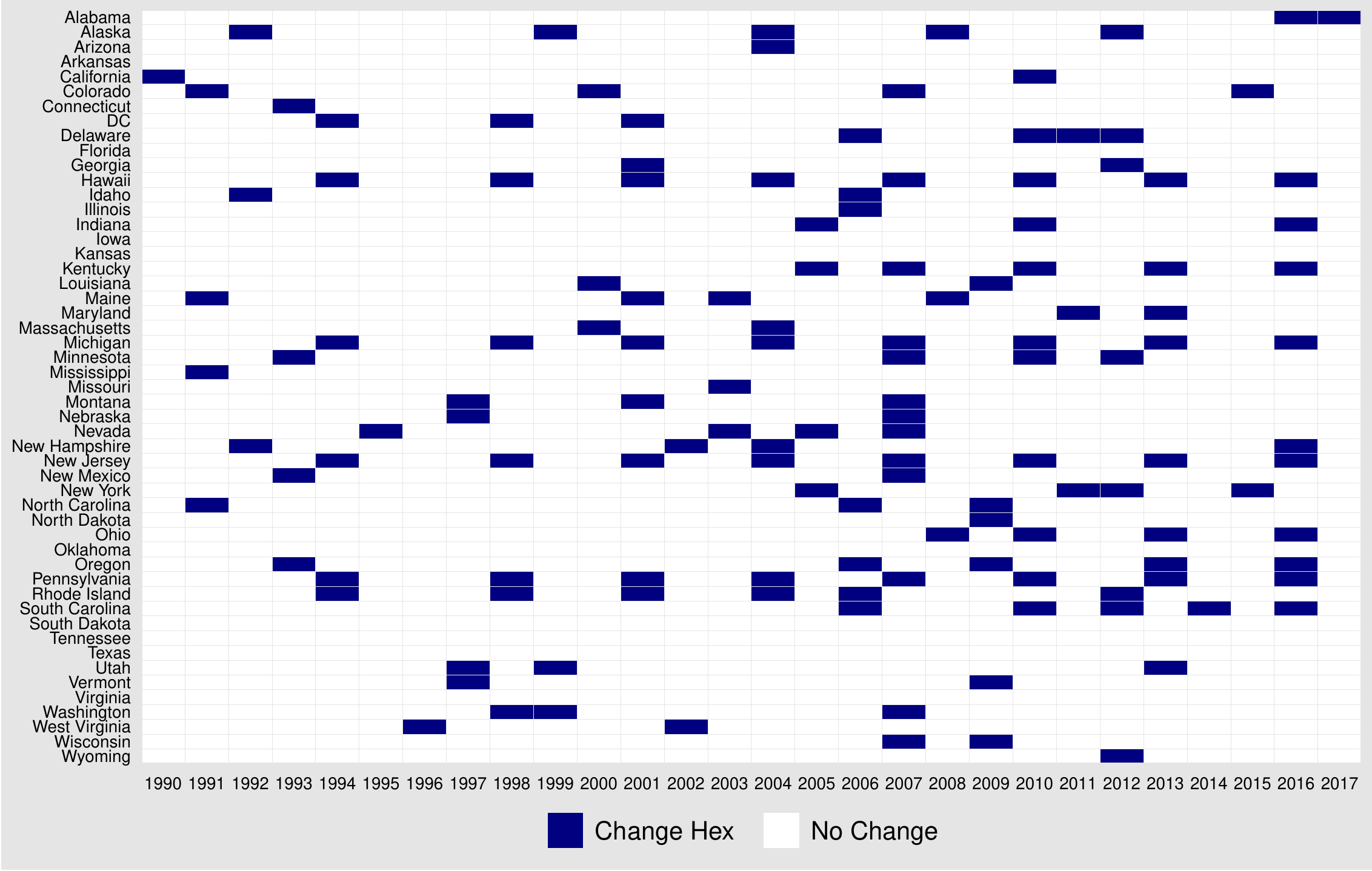}
	\begin{minipage}{0.7\textwidth}
		{\footnotesize 
			\textit{Note}. This figure presents all state changes in homestead exemption laws from 
1990-2017. This dataset includes all 50 states and Washington DC. 
In 2011, New York became the first state to offer county level exemptions for three groups of counties. 
Beginning April 2012, the New York homestead exemptions will be updated in April every 3 years to keep 
pace with inflation as measured by the New York-Newark-Jersey City CPI-U. 
The data was collected from other authors, bankruptcy guidebooks, and statutes as described in the paper.			  
			\par}
	\end{minipage}
\end{figure}

\newpage
\subsection{Homestead Exemption Levels}

\newpage
\subsection{Bankruptcies and Foreclosures}
\begin{figure}[!ht]
	\centering
	\small \caption{Bankruptcies and Foreclosures} 
	\label{FBankruptcies}
	\includegraphics[scale=.67]{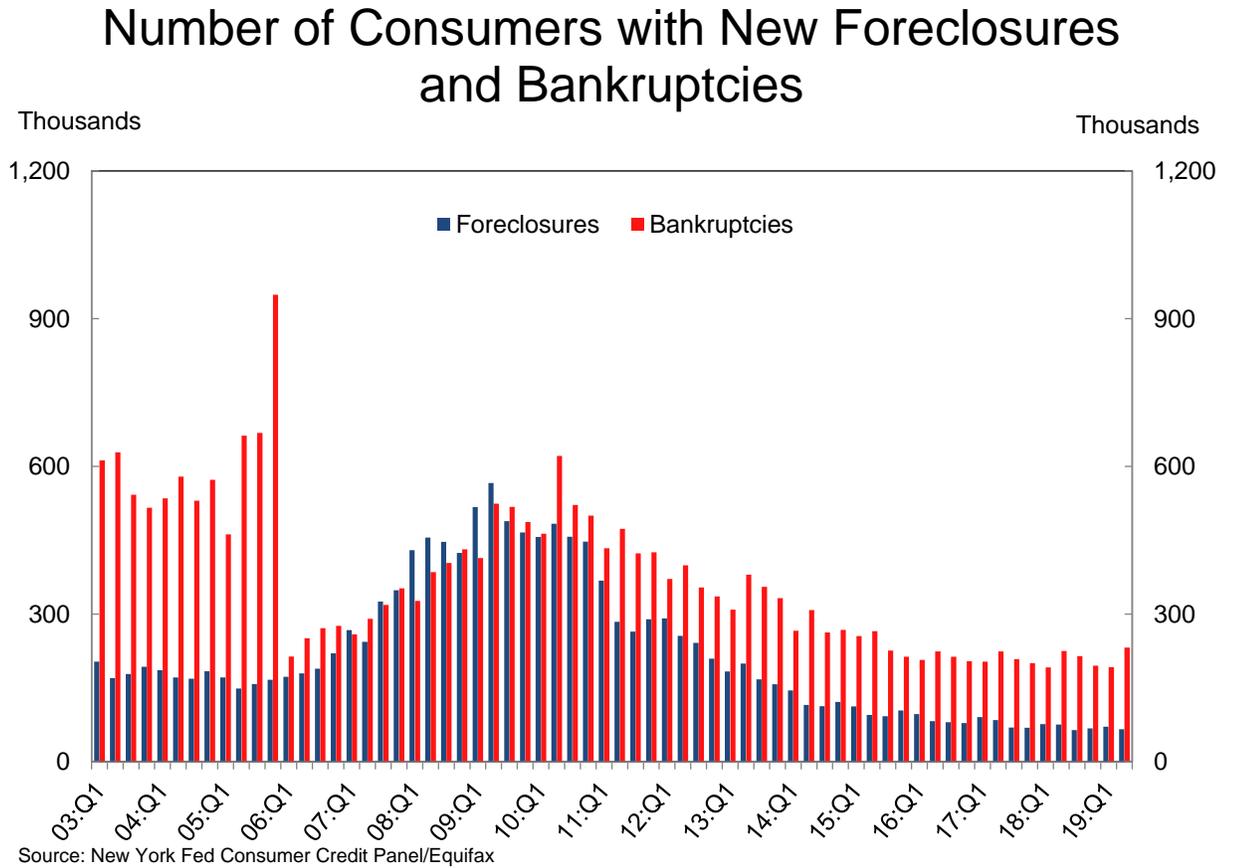}
	\begin{minipage}{0.7\textwidth}
		{\footnotesize 
			\textit{Note}. This figure plots the number of consumers with new bankruptcies and foreclosures
per quarter in the US. 
The data is from the New York Fed Consumer Credit Panel/Equifax.

\url{https://www.newyorkfed.org/microeconomics/hhdc/background.html}			
			\par}
	\end{minipage}
\end{figure}

\newpage
\subsection{Pre-Trends}

\newpage
\subsection{Supply Elasticity Theory}

\newpage
\section{Appendix: Tables}
\subsection{Descriptive Statistics}
\begin{table}[!htb]
	\centering
	\small \caption{Descriptive Statistics}
	\label{T1}
	{\scriptsize 	
\begin{center}
\begin{tabular}{lcccccc}
\hline \noalign{\smallskip}Variable & N & Min & Median & Mean & Max & Freq Change ($\%$)\\
\noalign{\smallskip}\hline \noalign{\smallskip} $ H_{i,1989}$ & 51 & 0.00 & 16,000.00 & 109,373.00 & 550,000.00 & \\
 $ H_{i,2005}$ & 51 & 0.00 & 50,000.00 & 151,559.00 & 550,000.00 & \\
 $ H_{i,2017}$ & 51 & 10,000.00 & 100,000.00 & 197,513.00 & 550,000.00 & \\
 $ \Delta H_{i,t}$ if $>0$ & 138 & 400.00 & 15,000.00 & 35,110.00 & 517,700.00 & 9.73\\
 $ \Delta H_{i,t}$ if $<0$ & 1 & -350,000.00 & -350,000.00 & -350,000.00 & -350,000.00 & \\
 $\% \Delta H_{i,t}$ if $>0$ & 136 & 1.00 & 33.30 & 84.50 & 1,603.00 & \\
\hline  $ NH_{i,1989}$ & 51 & 0.00 & 9,100.00 & 9,327.00 & 40,000.00 & \\
 $ NH_{i,2005}$ & 51 & 0.00 & 18,200.00 & 18,476.00 & 60,000.00 & \\
 $ NH_{i,2017}$ & 51 & 5,000.00 & 28,400.00 & 24,695.00 & 60,000.00 & 12.68\\
 $ \Delta NH_{i,t}$ if $>0$ & 180 & 200.00 & 2,000.00 & 4,404.00 & 30,000.00 & \\
 $ \Delta NH_{i,t}$ if $<0$ & 1 & -9,000.00 & -9,000.00 & -9,000.00 & -9,000.00 & \\
 $\% \Delta NH_{i,t}$ if $>0$ & 179 & 0.94 & 8.04 & 64.50 & 2,575.00 & \\
\noalign{\smallskip}\hline\end{tabular}\\
\end{center}
}
	\medskip
	\begin{minipage}{1\textwidth}
		{\footnotesize 
			\textit{Note}. 
This table reports descriptive statistics summarizing the homestead and non-homestead exemption
in the US between 1989-2017. 
The bankruptcy law data was collected from other authors, bankruptcy guidebooks, and statutes as described in the paper.	
$H_{i,t}$ is the maximum homestead exemption in census tract $i$ in year $t$. 
Similarly, $NH_{i,t}$ is the maximum non-homestead exemption (vehicle and wildcard) in census tract $i$ in year $t$. 
The percent change is undefined for two changes in $H_{i,t}$ (Delaware 2006, Maryland 2011)
which followed zero homestead exemption levels. 
The percent change is undefined for one such change in $NH_{i,t}$ (Delaware 2006).
The final column gives the frequency for \textit{all} (positive and negative) changes of each type.
Minnesota had a negative change in $H_{i,t}$ in 1993 
and  
Louisiana had a negative change in $NH_{i,t}$ in 2003.
			\par}
	\end{minipage}
\end{table}

\newpage
\subsection{Main Estimates}
\begin{table}[!htb]
	\centering
	\small \caption{Impact of Changes in Homestead Exemptions on Real House Price Growth}
	\label{US_T1}
	{\scriptsize 	
\begin{tabular}{lcccc} \hline
 & (1) & (2) & (3) & (4) \\
VARIABLES &  &  &  &  \\ \hline
 &  &  &  &  \\
$1\{\Delta H>0\}$ & 0.730*** &  &  &  \\
 & (0.226) &  &  &  \\
$1\{\Delta H>0\} \times \mbox{Pre-BAPCPA}$ &  & 1.066** &  &  \\
 &  & (0.418) &  &  \\
$1\{\Delta H>0\} \times \mbox{Post-BAPCPA}$ &  & 0.488 &  &  \\
 &  & (0.332) &  &  \\
$1\{\Delta H \geq 50k \}$ &  &  & 1.815*** &  \\
 &  &  & (0.548) &  \\
$1\{\Delta H< 50k\}$ &  &  & 0.411* &  \\
 &  &  & (0.239) &  \\
$1\{\Delta \geq 50k\} \times \mbox{Pre-BAPCPA}$ &  &  &  & 3.043*** \\
 &  &  &  & (0.923) \\
$1\{\Delta H< 50k\} \times \mbox{Pre-BAPCPA}$ &  &  &  & 0.395 \\
 &  &  &  & (0.471) \\
$1\{\Delta \geq 50k\} \times \mbox{Post-BAPCPA}$ &  &  &  & 0.742 \\
 &  &  &  & (0.955) \\
$1\{\Delta H< 50k\} \times \mbox{Post-BAPCPA}$ &  &  &  & 0.400 \\
 &  &  &  & (0.313) \\
 &  &  &  &  \\
N & 1266056 & 1266056 & 1266056 & 1266056 \\
R2 & .339 & .34 & .34 & .341 \\
 std-err & state & state & state & state \\ \hline
\multicolumn{5}{c}{ Robust standard errors in parentheses} \\
\multicolumn{5}{c}{ *** p$<$0.01, ** p$<$0.05, * p$<$0.1} \\
\end{tabular}
}
	\medskip
	\begin{minipage}{1\textwidth}
		{\footnotesize 
			\textit{Note}. 
This table reports estimates of the impact of a change in homestead exemption  
on real house price growth. 
Each column reports a separate regression estimated at the census tract year level 
where the dependent variable is the annual percent change of the real house price index. 
All specifications include census tract and year fixed effects. 
%
Standard errors, clustered at the state level, are reported in parentheses.
%
%

$1\{\Delta H>0\}$ is an indicator equal to one if census tract $i$ had a rise in the homestead 
exemption that year.
$\mbox{Pre-BAPCPA}$ is an indicator equal to one for years up to and including $2005$.
$\mbox{Post-BAPCPA}$ is an indicator equal to one for years after $2005$.
$1\{\Delta H>=50k\}$ is an indicator equal to one if census tract $i$ had a rise in the homestead 
exemption of at least \$50,000 that year.
%
%
%
House price data is at the census-tract year level from the FHFA, deflated by the CPI-U.
The bankruptcy law data was collected from other authors, bankruptcy guidebooks, and statutes as described in the paper.			  
\par}
	\end{minipage}
\end{table}

\newpage
\subsection{Validation: Dynamic Estimates}
\begin{table}[!htb]
	\centering
	\small \caption{Impact of Changes in Homestead Exemptions on Real House Price Growth}
	\label{T2_Dyncov}
	{\scriptsize 	
\begin{tabular}{lcccc} \hline
 & (1) & (2) & (3) & (4) \\
VARIABLES &  &  &  &  \\ \hline
 &  &  &  &  \\
$ X_{t-3} $ & -0.346 & -0.765 & -1.399 & -1.674 \\
 & (0.411) & (0.560) & (1.063) & (1.607) \\
$ X_{t-2} $ & -0.058 & 0.022 & -0.751 & 0.258 \\
 & (0.357) & (0.395) & (1.055) & (1.021) \\
$ X_{t} $ & 1.045*** & 1.416*** & 2.201*** & 3.415*** \\
 & (0.293) & (0.523) & (0.625) & (1.152) \\
$ X_{t+1} $ & 0.748 & 1.074 & 2.913*** & 3.758*** \\
 & (0.465) & (0.656) & (0.808) & (1.219) \\
$ X_{t+2} $ & 0.356 & 1.025* & 2.381*** & 2.163** \\
 & (0.457) & (0.561) & (0.805) & (0.958) \\
$ X_{t+3} $ & 0.016 & 0.164 & 1.547*** & 0.590 \\
 & (0.352) & (0.423) & (0.507) & (1.122) \\
 &  &  &  &  \\
X & $ 1 \{ \Delta H \gaz 0 \} $ & $ 1 \{ \Delta H \gaz 0 \}\times $ Pre-BAPCPA & $ 1 \{ \Delta H \geq 50k \} $ & $ 1 \{ \Delta H \geq 50k \}\times $ Pre-BAPCPA \\
N & 1012386 & 1012386 & 1012386 & 1012386 \\
R2 & .378 & .378 & .384 & .381 \\
 std-err & state & state & state & state \\ \hline
\multicolumn{5}{c}{ Robust standard errors in parentheses} \\
\multicolumn{5}{c}{ *** p$<$0.01, ** p$<$0.05, * p$<$0.1} \\
\end{tabular}
}
	\medskip
	\begin{minipage}{1\textwidth}
		{\footnotesize 
			\textit{Note}. 
This table reports estimates of the impact of a change in homestead exemption  
on real house price growth. 
Each column reports a separate regression estimated at the census tract year level 
where the dependent variable is the annual percent change of the real house price index. 
All specifications include census tract and year fixed effects. 
%
Standard errors, clustered at the state level, are reported in parentheses.
%
%
$1\{\Delta H>0\}$ is an indicator equal to one if census tract $i$ had a rise in the homestead 
exemption that year.
$\mbox{Pre-BAPCPA}$ is an indicator equal to one for years up to and including $2005$.
$1\{\Delta H\geq 50k\}$ is an indicator equal to one if census tract $i$ had a rise in the homestead 
exemption of at least \$50,000 that year.
%
%
%
House price data is at the census-tract year level from the FHFA, deflated by the CPI-U.
The bankruptcy law data was collected from other authors, bankruptcy guidebooks, and statutes as described in the paper.			  
			\par}
	\end{minipage}
\end{table}

\newpage
\subsection{Heterogeneity Analysis}
\begin{table}[!htb]
	\centering
	\small \caption{Heterogeneity in the Impact of Changes in Homestead Exemptions on Real House Price Growth}
	\label{THTE3}
	{\scriptsize 	
\begin{tabular}{lcccccccc} \hline
 & (1) & (2) & (3) & (4) & (5) & (6) & (7) & (8) \\
VARIABLES &  &  &  &  &  &  &  &  \\ \hline
 &  &  &  &  &  &  &  &  \\
$1\{\Delta H>0\}$ & 0.730*** & 1.677*** & 1.833*** & 0.494* & 0.185 & 0.311 & 2.711 & 0.750** \\
 & (0.226) & (0.538) & (0.653) & (0.284) & (0.706) & (0.374) & (2.977) & (0.281) \\
$1\{\Delta H<0\}$ & -0.311 &  &  &  &  &  &  &  \\
 & (0.375) &  &  &  &  &  &  &  \\
$1\{\Delta H>0\}\times\mbox{Elasticity}$ &  & -0.615*** &  &  &  &  &  &  \\
 &  & (0.217) &  &  &  &  &  &  \\
$1\{\Delta H>0\}\times\mbox{ur}_{t-1}$ &  &  & -0.195** &  &  &  &  &  \\
 &  &  & (0.094) &  &  &  &  &  \\
$1\{\Delta H>0\}\times\mbox{pop}_{t-1}$ &  &  &  & 0.002 &  &  &  &  \\
 &  &  &  & (0.003) &  &  &  &  \\
$1\{\Delta H>0\}\times\mbox{rincpc}_{t-1}$ &  &  &  &  & 0.026 &  &  &  \\
 &  &  &  &  & (0.034) &  &  &  \\
$1\{\Delta H>0\}\times\mbox{rzhvi}_{t-1}$ &  &  &  &  &  & 0.000 &  &  \\
 &  &  &  &  &  & (0.000) &  &  \\
$1\{\Delta H>0\}\times\mbox{hown}_{t-1}$ &  &  &  &  &  &  & -0.031 &  \\
 &  &  &  &  &  &  & (0.044) &  \\
$1\{\Delta H>0\}\times\mbox{permits1}_{t-1}$ &  &  &  &  &  &  &  & -0.000* \\
 &  &  &  &  &  &  &  & (0.000) \\
 &  &  &  &  &  &  &  &  \\
Observations & 1,266,056 & 1,012,642 & 1,258,840 & 1,266,530 & 1,266,530 & 932,422 & 1,282,709 & 1,256,147 \\
 R-squared & 0.339 & 0.374 & 0.324 & 0.330 & 0.317 & 0.382 & 0.329 & 0.341 \\ \hline
\multicolumn{9}{c}{ Robust standard errors in parentheses} \\
\multicolumn{9}{c}{ *** p$<$0.01, ** p$<$0.05, * p$<$0.1} \\
\end{tabular}
}
	\medskip
	\begin{minipage}{1\textwidth}
		{\footnotesize 
			\textit{Note}. 
This table reports estimates of the impact of a change in homestead exemption  
on real house price growth. 
Each column reports a separate regression estimated at the census tract year level 
where the dependent variable is the annual percent change of the real house price index. 
All specifications include census tract and year fixed effects. 
Standard errors, clustered at the state level, are reported in parentheses.
Each column reports a separate regression in which the treatment effect is allowed to vary based on 
seven measures of heterogeneity: supply elasticity, 
pre-law unemployment rate, population, real income per capita, real Zillow House Value Index, 
home ownership rate, and single family building permits.
%
%
%
House price data is at the census-tract year level from the FHFA, deflated by the CPI-U.
The bankruptcy law data was collected from other authors, bankruptcy guidebooks, and statutes as described in the paper.			  
Elasticity data are from (\cite{Saiz2010}), unemployment rates are from BLS, 
population, homeownership, and building permit data are from the Census, 
income per capita data are from the BEA, and House Value data are from Zillow. 
			\par}
	\end{minipage}
\end{table}

\newpage
\subsection{Mechanism: Alternative Outcome Variables}
\begin{table}[!htb]
	\centering
	\small \caption{Impact of Changes in Homestead Exemptions on Alternative Outcomes}
	\label{THTEY}
	\scalebox{.825}
	{\scriptsize 	
\begin{tabular}{lccccccccccc} \hline
 & (1) & (2) & (3) & (4) & (5) & (6) & (7) & (8) & (9) & (10) & (11) \\
VARIABLES & R\_HPG & ur & $\Delta\mbox{ur}$ & pop & $\Delta\mbox{pop}$ & incpc & $\Delta\mbox{incpc}$ & permits1 & $\Delta\mbox{permits1}$ & hown & $\Delta\mbox{hown}$ \\ \hline
 &  &  &  &  &  &  &  &  &  &  &  \\
$1\{\Delta H>0\}$ & 0.730*** & -0.018 & -0.051 & -0.950** & 0.056 & -0.284 & 0.091* & 362.960 & 32.041 & -0.402** & -0.136 \\
 & (0.226) & (0.085) & (0.051) & (0.454) & (0.049) & (0.265) & (0.050) & (269.946) & (65.669) & (0.156) & (0.102) \\
 &  &  &  &  &  &  &  &  &  &  &  \\
Observations & 1,266,056 & 1,355,759 & 1,300,184 & 1,362,513 & 1,307,877 & 1,362,513 & 1,307,877 & 1,352,999 & 1,297,701 & 1,380,113 & 1,324,795 \\
 R-squared & 0.339 & 0.810 & 0.649 & 0.991 & 0.769 & 0.919 & 0.330 & 0.770 & 0.121 & 0.941 & 0.162 \\ \hline
\multicolumn{12}{c}{ Robust standard errors in parentheses} \\
\multicolumn{12}{c}{ *** p$<$0.01, ** p$<$0.05, * p$<$0.1} \\
\end{tabular}
}
	\medskip
	\begin{minipage}{1\textwidth}
		{\footnotesize 
			\textit{Note}. 
This table reports estimates of the impact of a change in homestead exemption  
on five outcome variables and their differences. 
%
All specifications include census tract and year fixed effects. 
Standard errors, clustered at the state level, are reported in parentheses.
R\_HPG denotes real house price growth.
Each column reports a separate regression in which the outcome variable is the level and first difference 
of: 
unemployment rates, population, income per capita, single family building permits, and home ownership rates. 
%
%
%
House price data is at the census-tract year level from the FHFA, deflated by the CPI-U.
The bankruptcy law data was collected from other authors, bankruptcy guidebooks, and statutes as described in the paper.			  
Unemployment rates are from BLS, 
population, homeownership, and building permit data are from the Census, 
income per capita data are from the BEA. 
			\par}
	\end{minipage}
\end{table}

\newpage
\subsection{Predictors of Changes in Homestead Exemptions}
\begin{table}[!htb]
	\centering
	\small \caption{Predictors of Changes in Homestead Exemptions}
	\label{THTEp}
	\scalebox{.75}
	{\scriptsize 	
\begin{tabular}{lccccccccc} \hline
 & (1) & (2) & (3) & (4) & (5) & (6) & (7) & (8) & (9) \\
VARIABLES & $1\{\Delta H>0\}$ & $1\{\Delta H>0\}$ & $1\{\Delta H>0\}$ & 
$1\{\Delta H \geq 50k \}$ & $1\{\Delta H< 50k\}$ & $1\{\Delta \geq 50k\}$ & $1\{\Delta H< 50k\}$ 
& $1\{\Delta \geq 50k\}$ & $1\{\Delta H< 50k\}$ 
\\
  &   & $\times \mbox{Pre-BAPCPA}$ & $\times \mbox{Post-BAPCPA}$ & & 
   &  $\times \mbox{Pre-BAPCPA}$ & $\times \mbox{Pre-BAPCPA}$ & $\times\mbox{Post-BAPCPA}$ & $\times \mbox{Post-BAPCPA}$ 
\\ \hline
 &  &  &  &  &  &  &  &  &  \\
$ \mbox{RHPG}_{t-1}$ & 0.000 & 0.000 & 0.000 & 0.000 & -0.000 & 0.000 & -0.000 & 0.000 & 0.000 \\
 & (0.000) & (0.000) & (0.000) & (0.000) & (0.000) & (0.000) & (0.000) & (0.000) & (0.000) \\
$ \mbox{ur}_{t-1}$ & -0.002 & -0.005* & 0.003 & -0.003 & 0.001 & -0.003 & -0.003 & -0.000 & 0.003 \\
 & (0.005) & (0.003) & (0.004) & (0.003) & (0.004) & (0.002) & (0.003) & (0.002) & (0.003) \\
$ \mbox{pop}_{t-1}$ & 0.000 & 0.001** & -0.001* & 0.000 & -0.000 & 0.000 & 0.001** & 0.000 & -0.001** \\
 & (0.000) & (0.000) & (0.000) & (0.000) & (0.000) & (0.000) & (0.000) & (0.000) & (0.000) \\
$ \mbox{Rincpc}_{t-1}$ & -0.003 & 0.001 & -0.004** & -0.001 & -0.001 & -0.001* & 0.002* & -0.000 & -0.004** \\
 & (0.002) & (0.001) & (0.002) & (0.001) & (0.002) & (0.001) & (0.001) & (0.001) & (0.002) \\
$ \mbox{hown}_{t-1}$ & -0.007 & -0.000 & -0.006 & -0.002 & -0.004 & 0.000 & -0.001 & -0.003 & -0.003 \\
 & (0.005) & (0.004) & (0.004) & (0.003) & (0.004) & (0.002) & (0.003) & (0.002) & (0.004) \\
$ \mbox{H}_{i,t-1}$ & -0.007*** & -0.003*** & -0.004*** & -0.005*** & -0.003** & -0.003*** & -0.000 & -0.001*** & -0.003*** \\
 & (0.001) & (0.001) & (0.001) & (0.001) & (0.001) & (0.000) & (0.001) & (0.000) & (0.001) \\
$ \mbox{NH}_{i,t-1}$ & 0.000 & -0.001 & 0.001 & -0.001 & 0.001 & 0.000 & -0.001 & -0.002 & 0.002 \\
 & (0.002) & (0.002) & (0.002) & (0.001) & (0.003) & (0.001) & (0.002) & (0.001) & (0.002) \\
 &  &  &  &  &  &  &  &  &  \\
Observations & 1,217,887 & 1,217,887 & 1,217,887 & 1,217,887 & 1,217,887 & 1,217,887 & 1,217,887 & 1,217,887 & 1,217,887 \\
 R-squared & 0.279 & 0.295 & 0.308 & 0.290 & 0.282 & 0.317 & 0.289 & 0.238 & 0.299 \\ \hline
\multicolumn{10}{c}{ Robust standard errors in parentheses} \\
\multicolumn{10}{c}{ *** p$<$0.01, ** p$<$0.05, * p$<$0.1} \\
\end{tabular}
}
	\medskip
	\begin{minipage}{1\textwidth}
		{\footnotesize 
			\textit{Note}. 
This table reports estimates of the impact of several lagged predictors on indicators for
a change in homestead exemption.  
%
All specifications include census tract and year fixed effects. 
Standard errors, clustered at the state level, are reported in parentheses.
Each column reports a separate regression in which the outcome variable is an indicator equal to one if
a census tract experienced a rise in its homestead exemption.
The predictors are lagged: 
real house price growth, unemployment rates, population, real income per capita, 
homeownership rates, homestead exemption levels, and non-homestead exemption levels.
%
%
%
The bankruptcy law data was collected from other authors, bankruptcy guidebooks, and statutes as described in the paper.			  
House price data is at the census-tract year level from the FHFA, deflated by the CPI-U.
Unemployment rates are from BLS, 
population, homeownership, and building permit data are from the Census, 
income per capita data are from the BEA. 
			\par}
	\end{minipage}
\end{table}

\newpage
\section*{For Online Publication: Robustness Appendix }
\subsection{Main Estimates: Robustness}
\begin{table}[!htb]
	\centering
	\small \caption{Impact of Changes in Homestead Exemptions on Real House Price Growth}
	\label{US_T1rob}
	{\scriptsize 	
\begin{tabular}{lcccc} \hline
 & (1) & (2) & (3) & (4) \\
VARIABLES &  &  &  &  \\ \hline
 &  &  &  &  \\
$\Delta H$ & 1.219*** &  &  &  \\
 & (0.369) &  &  &  \\
$\Delta H\times \mbox{Pre-BAPCPA}$ &  & 2.164*** &  &  \\
 &  & (0.736) &  &  \\
$\Delta H\times \mbox{Post-BAPCPA}$ &  & 0.094 &  &  \\
 &  & (0.843) &  &  \\
$\Delta H\times 1\{\Delta H \geq 50k\}$   &  &  & 1.203*** &  \\
 &  &  & (0.373) &  \\
$\Delta H\times 1\{\Delta H < 50k\}$ &  &  & 1.617 &  \\
 &  &  & (1.948) &  \\
$\Delta H\times 1\{\Delta H \geq 50k\}\times \mbox{Pre-BAPCPA}$ &  &  &  & 2.182*** \\
 &  &  &  & (0.743) \\
$\Delta H\times 1\{\Delta H < 50k\}\times \mbox{Pre-BAPCPA}$ &  &  &  & 1.331 \\
 &  &  &  & (4.108) \\
$\Delta H\times 1\{\Delta H \geq 50k\}\times \mbox{Post-BAPCPA}$ &  &  &  & 0.008 \\
 &  &  &  & (0.894) \\
$\Delta H\times 1\{\Delta H < 50k\}\times \mbox{Post-BAPCPA}$ &  &  &  & 1.611 \\
 &  &  &  & (2.144) \\
 &  &  &  &  \\
Observations & 1,266,056 & 1,266,056 & 1,266,056 & 1,266,056 \\
 R-squared & 0.340 & 0.340 & 0.340 & 0.340 \\ \hline
\multicolumn{5}{c}{ Robust standard errors in parentheses} \\
\multicolumn{5}{c}{ *** p$<$0.01, ** p$<$0.05, * p$<$0.1} \\
\end{tabular}
}
	\medskip
	\begin{minipage}{1\textwidth}
		{\footnotesize 
			\textit{Note}. 
This table reports estimates of the impact of a change in homestead exemption  
on real house price growth. 
Each column reports a separate regression estimated at the census tract year level 
where the dependent variable is the annual percent change of the real house price index. 
All specifications include census tract and year fixed effects. 
%
Standard errors, clustered at the state level, are reported in parentheses.
%
%
$\Delta H$ is the change in the homestead exemption in a given year.
$1\{\Delta H>0\}$ is an indicator equal to one if census tract $i$ had a rise in the homestead 
exemption that year.
$\mbox{Pre-BAPCPA}$ is an indicator equal to one for years up to and including $2005$.
$\mbox{Post-BAPCPA}$ is an indicator equal to one for years after $2005$.
$1\{\Delta H>=50k\}$ is an indicator equal to one if census tract $i$ had a rise in the homestead 
exemption of at least \$50,000 that year.
%
%
%
House price data is at the census-tract year level from the FHFA, deflated by the CPI-U.
The bankruptcy law data was collected from other authors, bankruptcy guidebooks, and statutes as described in the paper.			  
			\par}
	\end{minipage}
\end{table}

\newpage
\subsection{Main Estimates: Robustness}
\begin{table}[!htb]
	\centering
	\small \caption{Impact of Changes in Homestead Exemptions on Real House Price Growth}
	\label{US_T1Dynrob}
	{\scriptsize 	
\begin{tabular}{lcccc} \hline
 & (1) & (2) & (3) & (4) \\
VARIABLES & $1\{\Delta H>0\}$ & $1\{\Delta H>0\} \times \mbox{Pre-BAPCPA}$ & $1\{\Delta H>= 50k \}$ & $1\{\Delta H>= 50k\} \times \mbox{Pre-BAPCPA}$ \\ \hline
 &  &  &  &  \\
$X_{t-3}$ & -0.322 & -0.038 & -0.400 & -0.007 \\
 & (1.199) & (1.540) & (1.259) & (1.561) \\
$X_{t-2}$ & -0.620 & 0.211 & -0.747 & 0.201 \\
 & (1.096) & (0.581) & (1.146) & (0.583) \\
$X_{t }$ & 1.396*** & 2.328*** & 1.384*** & 2.329*** \\
 & (0.379) & (0.787) & (0.382) & (0.791) \\
$X_{t+1}$ & 1.658*** & 2.677** & 1.853*** & 2.695** \\
 & (0.394) & (1.051) & (0.386) & (1.060) \\
$X_{t+2}$ & 1.220** & 1.236* & 1.504** & 1.127* \\
 & (0.593) & (0.645) & (0.623) & (0.593) \\
$X_{t+3}$ & 1.026** & 0.894** & 1.241*** & 0.868** \\
 & (0.451) & (0.364) & (0.463) & (0.351) \\
 &  &  &  &  \\
Observations & 988,952 & 988,952 & 988,952 & 988,952 \\
 R-squared & 0.382 & 0.382 & 0.383 & 0.382 \\ \hline
\multicolumn{5}{c}{ Robust standard errors in parentheses} \\
\multicolumn{5}{c}{ *** p$<$0.01, ** p$<$0.05, * p$<$0.1} \\
\end{tabular}
}
	\medskip
	\begin{minipage}{1\textwidth}
		{\footnotesize 
			\textit{Note}. 
This table reports estimates of the impact of a change in homestead exemption  
on real house price growth. 
Each column reports a separate regression estimated at the census tract year level 
where the dependent variable is the annual percent change of the real house price index. 
All specifications include census tract and year fixed effects. 
%
Standard errors, clustered at the state level, are reported in parentheses.
%
%
$\Delta H$ is the change in the homestead exemption in a given year.
$1\{\Delta H>0\}$ is an indicator equal to one if census tract $i$ had a rise in the homestead 
exemption that year.
$\mbox{Pre-BAPCPA}$ is an indicator equal to one for years up to and including $2005$.
$\mbox{Post-BAPCPA}$ is an indicator equal to one for years after $2005$.
$1\{\Delta H>=50k\}$ is an indicator equal to one if census tract $i$ had a rise in the homestead 
exemption of at least \$50,000 that year.
%
%
%
House price data is at the census-tract year level from the FHFA, deflated by the CPI-U.
The bankruptcy law data was collected from other authors, bankruptcy guidebooks, and statutes as described in the paper.			  
			\par}
	\end{minipage}
\end{table}

\newpage
\section*{For Online Publication: Appendix References }
\bibliographystyle{jf}

\end{appendices}
\end{document}